\documentclass[aps,twocolumn,showpacs]{revtex4}
\usepackage{psfig}

\newcommand{\ba}{\begin{eqnarray}}
\newcommand{\ea}{\end{eqnarray}}

\newcommand{\be}{\begin{equation}}
\newcommand{\ee}{\end{equation}}

\begin{document}

\title{
%
%
\[ \vspace{-2cm} \]
\noindent\hfill\hbox{\rm  } \vskip 1pt
\noindent\hfill\hbox{\rm SLAC-PUB-8999} \vskip 1pt
\noindent\hfill\hbox{\rm hep-th/0109201} \vskip 10pt
A Canonical Hamiltonian Derivation of Hawking Radiation
}

\author{Kirill Melnikov and Marvin Weinstein}
\affiliation{
Stanford Linear Accelerator Center\\
Stanford University, Stanford, CA 94309\\
E-mail: melnikov@slac.stanford.edu, niv@slac.stanford.edu}

\begin{abstract}
We present a derivation of Hawking radiation based on  
canonical quantization of a massless scalar field in the 
background of a Schwarzschild black hole using Lema\^itre
coordinates and show that in these coordinates the Hamiltonian
of the massless field is time-dependent. 
This result exhibits the non-static nature of the
problem and shows it is better to talk about the time dependence
of physical quantities rather than the existence of a time-independent
vacuum state for the massless field.
We then demonstrate the existence of Hawking radiation and show that
despite the fact that the flux looks thermal to an outside observer,
the time evolution of the massless field is unitary.
\end{abstract}


\maketitle

Hawking radiation \cite{hawking} is one of the most interesting phenomena
encountered in {\it pre-quantum\/} gravity.
It is a robust phenomenon in that it has been derived in different ways
(see e.g. \cite{unruh,jacobson,parikh})
and all derivations yield the same conclusion: 
a black hole of
mass $M$ emits (nearly) thermal radiation with a temperature 
$T_{\rm H} = 1/(8\pi G M)$.
To the best of our knowledge however, no derivation of Hawking
radiation discusses the problem within the framework of canonical quantization.
This is one reason why the issue of whether the time evolution of
the semi-classical theory is unitary has been a subject of debate.
In this paper we present a simple
canonical quantization procedure which leads to the usual
results within the framework of a unitary quantum theory.
Moreover, it enables us to compute the energy-momentum
tensor as a function of position for all times, allowing us,
as a matter of principle,  to give a self-consistent discussion 
of the back reaction problem.

The reason why it is not obvious that a canonical quantization
scheme is possible in Schwarzschild coordinates is because surfaces of fixed
{\it Schwarzschild time\/} are not globally space-like.
In Lema\^itre coordinates, however, surfaces of
constant time are space-like and extend from $r=0$ to $r = \infty$.
In this coordinate system one can canonically quantize a field on
a fixed time surface and then consider its time evolution
from that point on.
The immediate and most striking result of this approach 
is that the Hamiltionian of the massless field, computed in the
background of the Schwarzschild black hole, is time-dependent.  
From this point of view, there are no uniquely defined eigenstates 
(in particular the vacuum-state) of the the quantum mechanical
problem, in spite of the fact that we are dealing with 
a free field theory. It is this time dependence which 
is ultimately the origin of Hawking radiation.

The metric of the eternal black hole has the following form in Schwarzschild
coordinates:
\be
{\rm d} s^2 = -\left ( 1 - \frac{\alpha}{r} \right) {\rm d} t^2 
 + \left (1 - \frac{\alpha}{r}  \right )^{-1} {\rm d} r^2 
+ r^2 {\rm d} \Omega^2,
\ee
where $\alpha = 2GM$, $G$ is Newton's constant 
and $M$ is the mass of the black hole. Because the coefficient
of  $dr^2$ changes sign at $r = \alpha$ surfaces of
constant Schwarzschild time go from being space-like to time-like
at this point. This does not occur in Lema\^itre coordinates,
$\lambda,\eta$, which are related to $t,r$ by
\be
\lambda = t + 2 \sqrt{r \alpha} 
+ \alpha 
\ln \left | \frac{\sqrt{r}-\sqrt{\alpha}}{\sqrt{r}+\sqrt{\alpha}} \right |,
~~r = \alpha^{1/3}
\left [ \frac{3}{2} \left ( \eta - \lambda \right ) \right ]^{2/3},
\ee
In these coordinates the metric becomes:
\be
{\rm d} s^2  = -{\rm d} \lambda^2 + \frac{\alpha}{r} {\rm d}\eta^2 
+r^2 {\rm d} \Omega^2.
\ee
and from this form we see that surfaces of constant 
$\lambda$ are everywhere space-like.  Thus, these surfaces can be used to
carry out the canonical quantization of the field theory.
(Note, since $r$ is a function of 
$\lambda$, the metric is explicitly $\lambda$ or time-dependent
and this translates into the time-dependence of the Hamiltonian.)

The action of a minimally coupled massless scalar field in the background 
of the Schwarzschild black hole is
\be
S = \frac{1}{2} \int {\rm d}^4 x \sqrt{-g} g^{ij} 
\partial_i \Phi \partial_j \Phi.
\ee
Since this action is rotationally invariant we can
discuss  one angular momentum mode at a time.
From now on we  will focus on the $S$-wave component of the 
field and define $\phi_0 =  \Phi/\sqrt{4\pi}$.
In Lema\^itre coordinates standard manipulations lead to 
the Hamiltonian
\be
{\cal H} = \frac{1}{2} \int \limits_{\lambda}^{\infty} {\rm d} \eta 
\left \{ \frac {2\pi_0^2}{3\alpha (\eta-\lambda)} + 
\frac{3}{2} r (\eta - \lambda) 
\left ( \frac {\partial \phi_0}{\partial \eta} \right )^2 \right \},
\ee
where $\pi_0$ is the canonical momentum conjugate to $\phi_0$; i.e., 
$\pi_0 = \alpha (\eta - \lambda) \partial \phi_0/\partial \lambda$. 
As always,  $\pi_0$ and $\phi_0$ are assumed to satisfy the equal time 
commutation relation
$
[\pi_0(\lambda,\eta),\phi_0(\lambda,\eta')] = -i \delta(\eta - \eta').
$

The dynamics of the field $\phi_0$ and its canonical momentum $\pi_0$ 
are governed by the Hamilton equations of motion,
which are equivalent to the statement
\be
\alpha \frac {\partial }{\partial \lambda } \left [ (\eta - \lambda) 
\frac {\partial \phi_0}{\partial \lambda} \right ] 
- \frac {\partial }{\partial \eta } 
\left [ (\eta - \lambda) r \frac {\partial \phi_0}{\partial \eta} \right ]
 = 0.
\label{waveeq}
\ee
If we choose the initial space-like surface to correspond to
$\lambda =0$, so that $\eta = 2/3~r \sqrt{r/\alpha}$ and examine 
the $\lambda=0$ Hamiltonian, we see that it is useful to rescale
the field and its momentum as follows:
$\phi_0 = \phi_1/r$, $\pi_0 = \sqrt{r \alpha} \pi_1$.  After this rescaling
$\pi_1$ and $\phi_1$ satisfy the commutation relation:
$
[\pi_1(r),\phi_1(r')] = -i \delta(r - r'),
$
and the $\lambda=0$ Hamiltonian takes the simple form
\be
{\cal H}_{\lambda=0}  = \frac{1}{2} 
\int \limits_{0}^{\infty} {\rm d} r \left \{
\pi_1^2 + r^2 \left ( \frac {\partial }{\partial r} \frac {\phi_1}{r} 
\right )^2 \right \}.
\ee
Clearly, this is just the Hamiltonian for the radial mode of a free
scalar field and it is easily solved. As usual, self-adjointness
requires that we impose the boundary condition
$\phi_1(r=0)=0$, which implies that $\phi_1$ and $\pi_1$ can be written as
\ba
&& \phi_1(r) = \int \limits_{0}^{\infty}
\frac{{\rm d}\omega}{\sqrt{\pi \omega} } 
\sin(\omega r) \left ( a^+_{\omega} + a_{\omega} \right );
\nonumber \\
&& \pi_1 (r) = i \int \limits_{0}^{\infty}
\frac{\omega {\rm d}\omega}{\sqrt{\pi \omega} } 
\sin(\omega r)
 \left ( a^+_{\omega} - a_{\omega} \right ),
\label{eq10}
\ea
where $a^+_\omega,a_{\omega}$ are the usual creation and annihilation 
operators $[a_\omega,a^+_{\omega'}] = 2\pi \delta(\omega-\omega')$. The 
$\lambda=0$ vacuum state is defined by $a_\omega |0\rangle = 0$. 

For our purposes it will be sufficient to assume that at $\lambda=0$ 
the quantum field is in its vacuum state \cite{fen}. 
To study the time
dependence of the theory it is best to work in the Heisenberg
representation where the operators are functions of time and
the quantum state does not change. 
The time evolution of the Heisenberg fields are given by the wave equation 
Eq.(\ref{waveeq}) supplemented by the condition that $\phi_0$ and 
$\pi_0 \sim \partial \phi_0(\eta)/\partial \lambda$ reduce to their initial
values on the $\lambda=0$ surface.
While this is a complicated equation it will be sufficient, for our purposes,
to analyze it in the geometric optics approximation.  Eventually all formulas
for physical quantities will be expressed geometrically, so they will not
depend upon the coordinate system. To explain the geometric optics
approximation however, it is simplest to work in
Painleve coordinates; i.e.,  $\lambda,r$, since
the dependence of the solution on $\lambda$ and $r$ factorizes in these
coordinates.

If we look for solutions of the form
$\phi_0 = r^{-1} e^{i \omega \lambda} f_\omega(r)$, it 
is easy to see that, for large $\omega$,
$f_\omega(r)$ can be written  as
\ba
&& \ln f_\omega(r) = i\omega S_{1,2}(r) + {\cal O}(\omega^{-1}),
\nonumber \\
&&S_{1,2}(r) = \pm r -2 \sqrt{r \alpha} 
\pm 2 \alpha \ln |\sqrt{r/\alpha} \pm 1|.
\label{solutions}
\ea
Although the assumption of large $\omega$ may seem unjustified,
these two solutions do provide a very good approximation to the 
exact solutions both at $r/\alpha \to 1$ and $r \to \infty$, which are 
the two regions that are of primary interest to us. (Note, however, that these 
solutions are invalid for $r \to 0$, a point we will have more to say about 
below.) Examination of these solutions shows that they are characterized
by the fact that they are constant along ingoing or outgoing 
null geodesics.  Abstracting this fact we define the geometric optics
approximation as the assumption that the time evolution of the field $\phi_0$
is given by
\be
\phi_0(\lambda,r) = \frac {1}{r} \left [ \tilde \phi_1(\lambda + S_1(r)) 
 + \tilde \phi_2(\lambda + S_2(r)) \right ],
\label{e10}
\ee
where $\tilde \phi_1$ and $\tilde \phi_2$ are two arbitrary functions 
that can be 
determined from the requirement that at $\lambda=0$ they can be written
in terms of the field $\phi_0$ and its canonical momentum $\pi_0$.

Imposing this requirement to determine the two arbitrary functions 
$\tilde \phi_{1,2}(S_{1,2}(r)) = f_{1,2}(r)$ we find
\be
\phi_0(\lambda,r) = \frac{1}{r} \left \{ f_1(x_1) + f_2(x_2) \right \},
\ee
where 
\be
f_{1,2}(x) = \frac {1}{2} 
\int \limits_{0}^{x} {\rm d} \xi \left [ \phi_1'(\xi) 
\pm \pi_1(\xi) \mp \frac{\alpha^{1/2} \phi_1(\xi)}{\xi^{3/2}} \right ],
\label{funcf}
\ee
$\phi_1'= {\rm d}\phi_1/{\rm d}\xi$
and
$
S_{1,2}(x_{1,2}) = \lambda + S_{1,2}(r).
$
Now, since the field $\phi_1$ and the momentum $\pi_1$ are expressed through 
creation and annihilation operators defined at $\lambda=0$,
the above set of equations allows us to compute any Green's function 
of the field $\phi_0$ at any later time. 

A quantity which is of special interest is the expectation 
value of the $\lambda,\eta$ component of the energy-momentum tensor, since
it appears on the right-hand side of the Einstein equations. If
the integral of this quantity over a surface of fixed $r$ approaches a constant
as $r \rightarrow \infty$, this implies a
constant flux of outgoing energy, i.e., Hawking radiation. 
We wish to emphasize that the technique described below permits us to calculate
$T_{\lambda \eta}$ for any finite values of $\lambda$ and $\eta$ but for the
purposes of this Letter we will concentrate on the energy flux through an 
infinitely large sphere.

In Lema\^itre coordinates the total energy flux 
through an infinitely large sphere is given by
\be
{\cal J} = \lim_{\eta \to \infty} \int {\rm d} \phi {\rm d} \theta  
\sqrt{-g}  g^{\lambda \lambda} g^{\eta \eta} 
\frac{\langle T_{\lambda \eta} \rangle}{4\pi}
 =-\lim_{\eta \to \infty} \frac{r^{5/2}}{\alpha^{1/2}}
\langle T_{\lambda \eta} \rangle.
\label{flux}
\ee
where $\langle T_{\lambda \eta} \rangle$ is the vacuum expectation 
value of the off-diagonal component of the stress-energy tensor
of the field $\phi_0$ (a normalization factor of $(4\pi)^{-1}$
is introduced since $\phi_0$ denotes the $S$-wave component 
of the massless field).
Using the explicit expression for the time evolution of the field $\phi_0$,
Eq.(\ref{e10}), it is a straightforward matter to compute this flux.
In order to do it we need to regularize intermediate expressions
and we choose to regulate $T_{\lambda \eta}$ by point-splitting.
To this end we introduce $\sigma = \lambda_2 - \lambda_1$ and define
\be
\langle T_{\lambda \eta} \rangle = \lim_{\sigma \to 0} \frac{1}{2} 
\langle 0 | 
\left \{ \frac{\partial \phi_0 (\lambda_1,\eta)}{\partial \lambda }, 
\frac{\partial \phi_0(\lambda_2,\eta) }{\partial \eta}  \right \} |0 \rangle,
\label{fluxdef}
\ee
where $\{A,B\}$  denotes the anti-commutator of the fields.

The calculation of this quantity for large values of 
$\lambda$ and $\eta$ is simplest in  Painleve coordinates.
Since the value of $\eta$ is fixed, 
the equation $\lambda_2 - \lambda_1 = \sigma$ implies
that $r_1$ and $r_2$ are related by
\be
r_2 = r_1 - \frac{\sqrt{\alpha}\sigma}{r^{1/2}}
-\frac{\alpha \sigma^2}{4r_1^2}
-\frac{1}{6} \frac{\alpha^{3/2} \sigma^3}{r_1^{7/2}} + {\cal O}(\sigma^4).
\ee
Taking the derivatives in Eq.(\ref{fluxdef}) and considering  
the limit $r_{1,2} \to \infty$, we derive the following 
expression for the flux
\ba
&& {\cal J}= -\frac{1}{2} \lim_{\sigma \to 0} 
\left [ \langle \left \{
f_1'(x_1),f_1'(y_1) 
\right \} \rangle
W_1(r_2,r_1,x_1,y_1)
\right.
\nonumber \\
&& \left. +
\langle \left \{
f_2'(x_2),f_2'(y_2)
\right \} \rangle
W_2(r_2,r_1,x_2,y_2)
\right ],
\label{e16}
\ea
where the functions $f_{1,2}$ are defined in Eq.(\ref{funcf}),
$$
W_i(r_2,r_1,x,y) = \frac{S_i'(r_2)}{S_i'(x)S_i'(y)} 
\left (1 - \frac{\sqrt{\alpha} S_i'(r_1)}{\sqrt{r_1}} \right),
$$
and 
$x_{1,2}$, $y_{1,2}$ are defined through the set of equations:
\be
\lambda_{1} +S_1(r_1) = S_1(x_1),~~~\lambda_2 + S_1(r_2) = S_1(y_1),
\label{har}
\ee
and the same for $x_2,y_2$ with the change $S_1 \to S_2$.
Given these equations it is easy to derive the relation
between $y_{1,2}$ and $x_{1,2}$ as a power series expansion in $\sigma$.
It is clear from Eq.(\ref{har}) that the points $x_1,x_2$ are those
points on the $\lambda=0$ surface from which the null-geodesics  
which wind up at the point $\lambda,\eta$ must start out. From the
specific form of the functions $S_{1,2}$ we see that in the limit of
large $\lambda$ and large $\eta$, the limiting values for these points
are $x_1 \to \infty$, $x_2/\alpha \to 1$.

Let us now compute the anticommutator 
$\langle \{f_1'(x_1),f_1'(y_1)\} \rangle$.
Starting from Eq.(\ref{funcf}) we derive
\ba
 \langle \{f_1'(x_1),f_1'(y_1)\} \rangle
= && \frac{1}{4} \left [ 
\langle
\{\pi_1(x_1),\pi_1(y_1) \}\rangle
\right.  \\
&& \left. + \langle \left \{\phi_1'(x_1),
\phi_1'(y_1) \right\}\rangle 
+...\right],
\nonumber 
\ea
where the dots represent the terms which do not contribute to the flux in 
$r_1 \to \infty$ limit. The anticommutators in the above formula 
are computed using Eq.(\ref{eq10}) and  we find the energy flux
\ba
&& \langle \left \{\phi_1'(x_1),
\phi_1'(y_1) \right\} \rangle 
=
 \frac {-2}{\pi} \frac {(x_2^2+x_1^2)}{(x_1^2-x_2^2)^2};
\nonumber \\
&& \langle \{\pi_1(x_1),\pi_1(y_1) \} \rangle  = 
- \frac{4x_1y_1}{\pi (x_1^2 - x_2^2)^2}, 
\ea
so that the final result for the anticommutator reads
\be
\langle \{f_1'(x_1),f_1'(y_1)\} \rangle
 = -\frac{1}{2 \pi} \frac{1}{(x_1 - y_1)^2} + ...
\ee
Performing a similar calculation for the second term in Eq.(\ref{e16}) and 
substituting expansions of $y_{1,2}$ and $r_2$ in terms of $x_{1,2}$ and 
$r_1$, we find
\be
{\cal J} = \frac{1}{192 \pi \alpha^2} = \frac{\pi}{12} T_{\rm H}^2,
\ee
where we have introduced Hawking temperature $T_{\rm H} = 1/(8\pi GM)$.

We should make two comments concerning this result. First, the 
energy flux at large distances, we have just derived,  
is finite. This is in accord 
with the expectations regarding  possible divergences in the 
stress-energy  tensor computed 
in a gravitational background \cite{birel}. 
Quite generally, the result 
of such a calculation is divergent and the divergences are usually 
removed by appropriate renormalization. The allowed 
counterterms have a restricted form and include  
renormalization of the cosmological and Newtonian constants, 
as well as some other terms that do not appear in  Einstein's
action. It turns out that {\it none} of allowed  counterterms 
can  renormalize the {\it off-diagonal} element of the energy momentum
tensor in the gravitational background of a Schwarzschild black hole, 
both in Schwarzschild and Lema\^itre coordinates.
Therefore the result for the energy flux should come out finite and,
as we have seen,  it does.  The second comment concerns the calculation 
of the energy flux through the sphere of finite radius. It turns out that 
the result we obtain in that case
is not finite in that $\ln \sigma$ terms remain. 
We believe this to be due to the fact that we have
restricted attention to spherically symmetric configurations
and have not considered higher angular momenta.  This is 
supported by the fact that in the case of a two-dimensional
black hole, where higher angular momentum modes are absent, our result 
for the flux is finite for arbitrary values of $r$. We should emphasize 
that both the outgoing and the infalling fluxes are separately divergent
and only the sum of the two provides a finite, unambiguous answer.

To see that this time-independent flux of energy at large distances
corresponds to what one would expect from a body at a Hawking
temperature $T_{\rm H}$ it is necessary to weakly couple the
massless field to a detector \cite{unruh,dewitt79}
(which acts as a thermometer) located 
at some fixed Schwarzschild radius $r$.  In order to
make the computation more realistic we consider adiabatically 
switching on the  detector at some time $t=t_0$ for a finite amount of time 
$\delta$.  (Note, adiabaticity requires $E_{\rm typ} \delta \gg 1$,
where $E_{\rm typ}$ is typical detector level spacing.) To realize 
this situation we add an interaction term to the free field 
Lagrangian of the form $V_{\rm int} \sim e^{-(t-t_0)^2/(2\delta^2)} \phi_0(t,r) \hat M$,
where $\hat M$ is an operator which acts in the Hilbert space of detector 
eigenstates. It follows from second order perturbation theory that the probability
of exciting the detector to a state of energy $E$ is proportional to
the Fourier transform of the Green's function
of the massless field \cite{birel}:
\ba
&& {\cal P}(\Delta E) \sim |\langle E | M(0) | E_0 \rangle |^2
\int {\rm d} t {\rm d} t' e^{-i\Delta E (t-t') } \times
\nonumber \\
&& e^{-[(t-t_0)^2+(t'-t_0)^2]/(2\delta^2)} 
\langle  \phi_0(t,r) \phi_0(t',r)  \rangle,
\label{detex}
\ea
where $\Delta E = E-E_0$ and $E_0$ is the ground state energy of the detector. 
As in the calculation of the 
flux which we already described, the Green's function in Eq.(\ref{detex})
is computed using the evolution equation for the field $\phi_0$ which relates it 
to initial conditions for  $\phi_0$ and $\pi_0$  on the 
surface $\lambda=0$. It is convenient to define 
$x_{1,2}$ and $y_{1,2}$ as in Eq.(\ref{har}) and identify 
$r_1=r_2=r$, $\lambda_1 = \lambda_1(t,r)$ and 
$\lambda_2 = \lambda_2(t',r)$. As before, these are the points from which 
infalling and outgoing geodesics leave the $\lambda=0$ surface 
in order to reach the points $t,r$ and $t',r$. Now consider Eq.(\ref{detex})
for large values of $t_0$. Using the explicit form of the functions 
$S_{1,2}$ it is easy to find the following approximate solutions:
$ x_1 = t +r,~~y_1=t'+r,~~ 
x_2 = \alpha ( 1 + 2  e^{-(t-r)/(2\alpha)} ),~~
y_2 = \alpha  ( 1 + 2  e^{-(t'-r)/(2\alpha)} ),
$
with $t \sim t' \sim t_0$. It is then clear that, asymptotically,
$x_2 \to y_2 \to \alpha$ and $x_1 \to y_1 \to \infty$.
In this limit, the Green's function can be written as
\ba
&& \langle \phi_0(t,r) \phi_0(t',r) \rangle \approx
\frac{-1}{4\pi r^2} \left ( 
\ln|x_1 - y_1| + \ln|x_2 - y_2| 
\right . \nonumber \\
&& \left.
 + \frac{i \pi}{2} 
\left [ 
\kappa (x_1,y_1) + \kappa (y_2,x_2)
\right ]
+ c
\right ),
\label{gf}
\ea
where $\kappa(x,y) = \theta(x-y) - \theta(y-x) $ 
and 
$c$ is some constant.

It is instructive to consider the terms in 
Eq.(\ref{gf}) separately.  The constant does not contribute to 
${\cal P}(E)$ since it  yields a result proportional 
to $\exp(-\Delta E^2 \delta^2) \ll 1$.
The $\ln|x_1 - y_1|$ term and the terms described by the function 
$\kappa$ give simple, $\alpha$ independent, contributions that 
can be written as
\be
{\cal P}_1(\Delta E) \sim - \frac{\pi \delta}{\Delta E} 
+ {\cal O}((\delta \Delta E)^{-1}). 
\label{c1}
\ee
The important part of the final result 
comes from the second term in 
Eq.(\ref{gf}) which describes the radiation coming from the vicinity 
of the horizon. Appropriately shifting the integration 
variables we obtain
\ba
{\cal P}_2(\Delta E) && \sim
- \int {\rm d} t {\rm d} t' e^{-i\Delta E (t-t') }
e^{-[(t-t_0)^2+(t'-t_0)^2]/(2\delta^2)}
\nonumber \\
&& \times \ln|e^{-t/(2\alpha)}-e^{-t'/(2\alpha)}|.
\ea
If we then change the variables to $v=(t+t'),~u=(t-t')$ and neglect 
all the suppressed terms we arrive at
\be
{\cal P}_2(\Delta E)\sim \frac{2\pi \delta}{\Delta E}
\left [
\frac{1}{e^{\Delta E/T_{\rm H}} -1} 
+ \frac{1}{2} 
\right ].
\ee
Since the total probability is given by the sum of ${\cal P}_1$ and ${\cal P}_2$
we get the final result:
\ba
\frac{{\cal P}(\Delta E)}{\delta} \sim 
\frac{|\langle E | M(0) | E_0 \rangle |^2 }{\Delta E}
\times
\frac{1}{e^{\Delta E/T_{\rm H}} -1}.
\ea
The interpretation of this formula is straightforward.
If, at a large distance from the black hole, an observer switches on
a detector which interacts with the massless field for finite amount of time, 
then the energy levels of the detector get populated as if the detector was
in equilibrium with a thermal distribution of particles at a 
temperature $T_{\rm H}$.

In conclusion, in this Letter we have presented a canonical
Hamiltonian derivation of Hawking radiation for the case of a
Schwarzschild black hole.  In this formalism
the Hamiltonian of the system is perforce time-dependent and so
the discussion of the vacuum state of the massless field is
irrelevant to the long time behavior of the system.
By the very nature of our construction, the state of the system
is always a pure state and for this reason there are correlations 
of the fields  inside and outside the black hole horizon.  
We also explicitly demonstrated, in agreement with arguments that 
concern the general structure of the divergences in the energy-momentum 
tensor in the gravitational background, that the Hawking flux at
infinity is a uniquely defined finite quantity.
This opens up the possibility of identifying the time-dependent 
function that should be inserted into the right hand side of the 
Einstein equations to study the back reaction of the Hawking 
radiation on the geometry of the space-time in a self-consistent 
way. 

Clearly, this approach allows us to address more issues than we
can touch upon in a Letter; in particular, the question of
back-reaction.  Such issues will be discussed in detail in
a long paper which is in preparation.  It would be remiss of us, however,
to conclude without pointing out that there is 
a significant difference between the case of the two and 
four-dimensional
black hole.  Direct substitution shows that the solutions 
in Eq.(\ref{solutions})
are {\it exact} for the two-dimensional black hole.  This is not
true for the four-dimensional black hole where the geometric optics
approximation breaks down at $r \to 0$.  Thus, the
behavior of the quantum theory near $r \to 0$ is different
in the two cases.  This is important because study of both cases shows
that after a finite time some modes of the field no longer appear in
the time-dependent Hamiltonian and thus, their
time evolution becomes trivial.  It requires careful study of the
behavior of the system in order to understand whether or not
these modes stay decoupled as the hole evaporates and whether they store
any information. 

This research was supported by the DOE under grant number
DE-AC03-76SF00515.


\begin{thebibliography}{10}
\bibitem{fen}  If the initial state is not the vacuum state but 
any finite energy state of the instantaneous Hamiltonian, the 
finite energy will be radiated away in finite amount of time.
The structure of radiation at large later times is independent 
of the initial quantum state of the massless field.
\end{thebibliography}

\end{document}